\documentclass[12pt,a4paper,final]{iopart}
\expandafter\let\csname equation*\endcsname\relax

\expandafter\let\csname endequation*\endcsname\relax
\usepackage{graphicx}
\usepackage{support-caption}
\usepackage{subcaption}
\usepackage{blindtext}
\usepackage{amsmath}
\usepackage{amssymb}
\usepackage[titletoc,page]{appendix}
\usepackage{color}
\usepackage[utf8]{inputenc} % usually not needed (loaded by default)
\usepackage[T1]{fontenc}

\usepackage{etoolbox}
\usepackage{chngcntr}
\AtBeginEnvironment{appendices}{%
  \counterwithin{figure}{section}
  \counterwithin{table}{section}
}

\usepackage{hyperref} %[breaklinks=true,colorlinks=true,linkcolor=blue,urlcolor=blue,citecolor=blue]

\usepackage{lineno}
\newcommand{\diff}{{\rm d}}

\newcommand{\jose}[1]{{\color{black}{#1}}} 
\newcommand{\oyvind}[1]{{\color{black}{#1}}} 
\newcommand{\revoyvind}[1]{{\color{black}{#1}}} 
\newcommand{\spelloyvind}[1]{{\color{black}{#1}}}

\long\def\old#1{}

\begin{document}

\title[Charged Black Hole Mergers: Orbit Circularisation and Chirp Mass Bias]{Charged Black Hole Mergers: Orbit Circularisation and Chirp Mass Bias}

\author{Øyvind Christiansen}
\address{Institute of Theoretical Astrophysics\\University of Oslo\\Sem Sælands vei 13, 0371 Oslo}
\ead{oyvind.christiansen@astro.uio.no}

\author{Jose Beltr\'an Jim\'enez}
\address{Departamento de F{\'i}sica Fundamental and IUFFyM\\ Universidad de Salamanca\\
Plaza de la Merced s/n, E-37008, Salamanca (Spain).}
\ead{jose.beltran@usal.es}

\author{David F. Mota}
\address{Institute of Theoretical Astrophysics\\University of Oslo\\Sem Sælands vei 13, 0371 Oslo}
\eads{\mailto{d.f.mota@astro.uio.no}}

\begin{abstract}
We consider the inspiral of black holes carrying U(1) charge that is not electromagnetic, but corresponds to some dark sector. In the weak-field, low-velocity regime, the components follow Keplerian orbits. We investigate how the orbital parameters evolve for dipole-dominated emission and find that the orbit quickly circularises, though not as efficiently as for a gravitationally dominated emission. We then regard circular orbits, and look for modifications in the gravitational waveform from the components carrying small charges. Taking this into account we populate the waveform with simplified LIGO noise and put it through a matched filtering procedure where the template bank only consists of uncharged templates, focusing on the charges' effect on the chirp mass estimation. We find a consistent overestimation of the `generalised' chirp mass, and a possible over- and underestimation of the actual chirp mass.
Finally, we briefly consider the effect of such charges on hyperbolic encounters, finding again a bias arising from interpreting the generalised chirp mass as the actual chirp mass.
\end{abstract}

%Uncomment for PACS numbers title message
%\pacs{00.00, 20.00, 42.10}
% Keywords required only for MST, PB, PMB, PM, JOA, JOB? 
%\vspace{2pc}
%\noindent{\it Keywords}: Article preparation, IOP journals
% Uncomment for Submitted to journal title message
%\submitto{\JPA}
% Comment out if separate title page not required
\maketitle
%\newpage 

\tableofcontents

\markboth{}{}
\renewcommand{\thefootnote}{\fnsymbol{footnote}}

\section{Introduction}
    Nearly a century after the theoretical prediction of the existence of Gravitational Waves (GWs) by Einstein \cite{einstein_uber_1918}, the LIGO interferometers \cite{TheLIGOScientific:2014jea} culminated a long quest with the first ever direct detection of a GW \cite{ligo_scientific_collaboration_and_virgo_collaboration_observation_2016}. Although indirect evidence for their existence had already been provided by binary pulsars, e.g. the celebrated Hulse-Taylor pulsar and many subsequent ones (see \cite{Wex:2014nva} and references therein), the LIGO detection opened a new window to observe the universe with the potential to discover new astrophysical objects. The incorporation of the Virgo interferometer \cite{TheVirgo:2014hva} to the network increased the sensitivity to the GWs' polarisations and allowed a more precise positioning of the sources \cite{Abbott:2017oio}. Since then a substantial number of events have been observed by the LIGO/Virgo collaboration \cite{ligo_scientific_collaboration_and_virgo_collaboration_gwtc-1_2019}. A major breakthrough came about with the first detection of a merger of two neutron \revoyvind{stars} \cite{TheLIGOScientific:2017qsa}, whose electromagnetic signal could also be observed \cite{GBM:2017lvd}. This event marked the dawn of multi-messenger astronomy.
    
    %In their detection paper \cite{ligo_scientific_collaboration_and_virgo_collaboration_observation_2016}, almost a century after the theoretical prediction , the LIGO-Virgo collaboration (LVC) could finally declare to have detected gravitational waves. It was coined the dawn of multi-messenger astronomy, and we've since observed an increasing number of types of events through this brand new channel \cite{ligo_scientific_collaboration_and_virgo_collaboration_gwtc-1_2019}.
    
    GWs provide an entirely novel detection channel with excellent prospects to peer into the strong gravity regime and the high-energy physics involved in the mergers of Black Holes (BHs) and Neutron Stars (NSs), thus \spelloyvind{providing} new opportunities to discover physics beyond the standard models. However, clear signatures of new physics have not been observed so far and the predictions of General Relativity (GR) seem to be able to account for the detected signals
     \cite{the_ligo_scientific_collaboration_and_the_virgo_collaboration_tests_2019}, although the obtained typical mass ranges of the events as well as GW170729 has challenged some stellar mass black hole formation models \cite{ligo_scientific_and_virgo_collaboration_gw170104_2017,chatziioannou_properties_2019}. In any case, we still lack any robust evidence for the existence of more exotic, though plausible, objects like hairy black holes, black hole mimickers, boson stars, compact objects of modified gravity scenarios, etc. (see e.g. \cite{Cardoso:2019rvt} for a recent status report). 
    
   The catalogue paper of the LIGO/Virgo collaboration \cite{ligo_scientific_collaboration_and_virgo_collaboration_gwtc-1_2019} reports on an increase in significance of some events due to improvements made on the detector pipelines. It is also reported that, although the number of marginal events detected are not unlikely at the current thresholds, it is currently difficult to discern the astrophysical origin of any of them. Consequently, a potential reason for the non-discovery of exotic objects like the ones mentioned above could be that LIGO/Virgo's most sensitive detection pipelines, the matched filtering pipelines of PyCBC \cite{usman_pycbc_2016} and GstLAL \cite{sachdev_gstlal_2019}, require an accurate theoretical modelling of the astrophysical source with templates that would allow to reliably extract the signal out of the noise \cite{sathyaprakash_choice_1991}. These types of objects' theoretical predictions are not included in the template banks that are used in the pipelines, and so their signal might either
    go undetected or not be considered significant enough and become a marginal event. It could also be the case that some of these signals might actually have been detected, but misinterpreted in terms of the existing templates.

    The no-hair theorem of BHs in GR \cite{saa_new_1996} states that a BH is completely described by its mass, spin and \oyvind{electric and magnetic} charge. This is a non-trivial and paramount result for the physics of BHs since it allows to reduce any BH solution to just three numbers, in high contrast to e.g. a neutron star whose extremely complicated internal structure is required to understand its behaviour. Furthermore, the electric neutrality of the universe makes it so that actual BHs can only carry a negligible charge \cite{zajacek_electric_2019} and they can therefore be well-described in terms of only their mass and spin. For binary systems, this simple picture drawn by the no-hair theorem allows us to create a discretised grid for the binary's relatively low-dimensional parameter space on which numerical templates are constructed, densely enough to hopefully cover all of the true binary parameter space. However, this has the intrinsic limitation that exotic compact objects not complying with the no-hair theorem, or better described by solutions other than the Kerr metric, cannot be properly captured by the templates. It is also possible that the orbit has a significant residual eccentricity by the time it enters LIGO/Virgo sensitivity (the templates are obtained for the already circularised regime), but this would only happen under special circumstances \cite{abbott_search_2019} due to the orbits circularising from GW emission by the time of merging for initially large separation black hole binaries.
    
     The main purpose of this work is to explore some observational consequences of compact objects beyond those well-captured by the employed templates. We will however remain very conservative and only consider the case of charged BHs, whose existence do not require introducing any severe extensions to the standard model nor any modification of the gravitational interaction.
     As commented above, charged BHs are traditionally dismissed due to the electric neutrality of the universe on the relevant scales, although there are some suggested mechanisms through which an astrophysical BH could acquire some electric charge, though a very small amount \cite{zajacek_electric_2019,wald_black_1974,levin_black_2018,cohen_magnetospheres_1975,lee_electric_2001}. Although we are mainly interested in studying the potential effects of a population of charged BHs while remaining agnostic about their origin or formation process, we can give some well-motivated scenarios where such objects could be formed.
     
     One of such scenarios is a class of self-interacting Dark Matter (DM) models where the interaction is mediated by some {\it dark} photon (see e.g. \cite{cardoso_black_2016,de_rujula_charged_1990, plestid_new_2020} for similar motivations to consider charged BHs). After all, all the interactions in the standard model of particles (leaving gravity aside) are mediated by massless spin-1 gauge bosons and, after symmetry breaking, the photon remains as the relevant one on large (as compared to the corresponding Compton wavelengths) scales, so it is plausible that the dark sector \spelloyvind{also could} contain some dark U(1) gauge field. Obviously, some mechanism in the early universe should lead to a net dark charge so charged BHs will not be neutralised by accreating DM particles of the opposite charge\footnote{The BH could also be neutralised by Hawking radiation of charged particles, but this process is typically very slow and does not prevent the existence of long-lasting charged BHs\spelloyvind{,} provided they are sufficiently big. The potentially strong (dark) electric field near the horizon could also contribute to discharging the BH via the Schwinger effect. This depends on the specific details of the model like the available phase space, the mass of the products, the charges, the dark U(1) coupling constant, etc. We will assume that the discharge time-scale for this process is sufficiently long.}.
    % Still, we might also imagine some dark sector fractional electric or hidden charge \cite{cardoso_black_2016,de_rujula_charged_1990, plestid_new_2020}, whose carriers might have a matter/antimatter asymmetry caused by a similar mechanism to that of the baryon asymmetry \cite{canetti_matter_2012}. In that case the universe might not be neutral, 
    The formation of BHs from these DM particles (for instance if their generated spectrum presents high peaks on small enough scales) would then give rise to a population of charged BHs. On the other hand, these charged BHs could have been produced by some other unknown mechanism in the early universe, so they do not need to be related to DM at all, although their relation certainly substantiates the motivation. In any case, associating the electric charge not to that of electromagnetism, but to some dark (hidden) sector permits us to sidestep the commonly invoked objection towards charge-inclusion in black hole mergers. Let us emphasise one last time that our scenario, aside from better or worse theoretical motivations, intends to simply study some consequences of having a population of a less explored region in the solution space of GR, namely: that of Reissner-Nordström BHs \cite{reissner_uber_1916}, or Kerr-Newman \cite{newman_note_1965} in the rotating case.
    
%    \jose{There are also other motivations for looking at charged black holes. As has been shown in \cite{zhang_mergers_2016}, the study of electrically charged black hole mergers might potentially offer an explanation for the still mysterious fast radio and gamma ray bursts. Adding charged-templates to the template bank might realise their detection in the GW channel, which would offer insight into the behaviour of electromagnetism in extreme environments.} Another possibility is that the charged merger events are discovered but wrongly interpreted during parameter estimation --  in this article we investigate the effect of weakly charged black hole mergers on the matched filtering parameter estimation with only neutral merger templates in the template bank.}
    
    The article is structured in the following way: We start in Section \ref{model} by considering Keplerian orbits where the emission is dominated by the electromagnetic (EM, here meaning dark sector electromagnetic) charge dipole and find the time-evolution of the orbital parameters. In Section \ref{circularisation} we consider how quickly the orbit circularises as its semi-major axis decreases. In Section \ref{linearised} we move on to considering circular orbits with BHs emitting in both EM charge dipole and GW mass and EM charge\footnote{
    When the dipole is not suppressed relative to the charge quadrupole due to small charge to mass ratio difference, the Biot-Savart \oyvind{and relativistic} corrections to the dipole would have to be included to have the complete radiation at that order. For the complete 1PN Lagrangian see (3.1) of \cite{khalil_hairy_2018}.
    }
    quadrupole channels, but where the dipole power is much smaller than the quadrupole power. \revoyvind{Linearising in a quantity proportional to the square of the black holes' charge-to-mass ratio difference}, we find the resulting GW waveform which we populate with noise and use as a mock strain on which we perform the matched filtering parameter estimation for the chirp mass. We present our results in \ref{results}. In \ref{hyperbolic} we briefly consider the consequences of a similar analysis on hyperbolic trajectories.    \\
    %\newpage

\noindent
{\bf Notation:} Throughout this paper we will unless otherwise stated use geometrised, natural units: $G=c=\frac{1}{4\pi\epsilon_0} = 1$. We will denote the time-average of a quantity $A$ as $\langle A \rangle$. We use Einstein summation convention, and use Greek indices for vectors/tensors spanning space and time, while using Roman indices for purely spatial vectors/tensors. Although the charges are intended to be hidden and not electromagnetic, we will often refer to their quantities using `electric', `magnetic' or with the abbreviation EM.

\section{Modelling the Merger}
\label{model}

We are interested in a binary system of BHs that we model as point-like sources carrying both mass and charge, i.e. they are well-described by the following energy and charge densities

\begin{equation}
\rho_{m}(\vec{x})= \sum_{a=1,2} m_a \delta^{(3)}(\vec{x}-\vec{x}_a),\quad
\rho_{q}(\vec{x})= \sum_{a=1,2} q_\revoyvind{a} \delta^{(3)}(\vec{x}-\vec{x}_a),\label{densities}
\end{equation}

\noindent
with $m_\revoyvind{a}$ and $q_\revoyvind{a}$ the corresponding mass and charge of the two BHs and $\vec{x}_\revoyvind{a}$ their trajectories, \revoyvind{$a$ specifying which we are referring to}. We will work at the lowest order so that their internal structure \spelloyvind{is not} relevant; It will only come into play at higher post-Newtonian orders \cite{blanchet_gravitational_2014}.
This characterisation of the charged binary system is similar to that of \cite{peters_gravitational_1963} and \cite{liu_merger_2020}. As usual, it is convenient to work in the centre of mass system defined by

\begin{equation}
    m_1\vec{r}_1+m_2\vec{r}_2=0,
\end{equation}
\noindent
and introduce the relative position vector $\vec{R}\equiv \vec{r}_1-\vec{r}_2$ so \spelloyvind{that} we have the following relations

\begin{equation}
    \vec{r}_1 = \frac{m_2}{M}\vec{R},\quad
    \vec{r}_2 = -\frac{m_1}{M}\vec{R},
\end{equation}

\noindent
with $M = m_1 + m_2$. The charge\spelloyvind{s} of the BHs \jose{introduce} an additional electric force that gives rise to a correction to the usual gravitational potential of the system. However, since this extra potential shares the Newtonian $1/R$ form, we can easily incorporate its effect with the potential \revoyvind{energy}

\begin{equation}
V = -\Tilde{G}\frac{\mu M}{R}    ,
\end{equation}

\noindent
where $\mu = m_1 m_2/M$ is the reduced mass and we have introduced the effective Newton's constant\footnote{Notice that we are working in units of $G=1$ so restoring the factors of $G$ we would have $\Tilde{G}\rightarrow\Tilde{G}/G$.}

 \begin{equation}
 \Tilde{G} = 1 - \sigma_1\sigma_2,
\end{equation}

\noindent
with $\sigma_i \equiv q_i/m_i$ the charge-to-mass ratios. It is then immediate to obtain the corresponding Keplerian orbits that are characterised by

\begin{align}&
R = \frac{a(1-e^2)}{1+e\cos\psi},\quad
    a = \frac{\tilde G M \mu}{2 | E |},\\&
    x = R\cos\psi,\quad y = R\sin\psi, \label{cartesian}
\end{align}

\noindent
with $a$, $e$, \revoyvind{$E$} and $\psi$ the semi-major axis, eccentricity, \revoyvind{orbital energy} and the polar coordinate of the reduced mass 1-body problem \revoyvind{respectively}.

\subsection{Energy and Angular Momentum Emission}

Our assumption of weak-field, low velocity motion translates to only keeping the lowest terms of the multipole expansion for the radiation fields. For the electromagnetic field in the Coulomb gauge (with $A^0=0$) we then need to solve the \revoyvind{flat space} wave equations

\begin{equation}
    \square A^i = j^i,
\end{equation}

\noindent
where $j^i$ is the corresponding conserved current owing to the charges of the BHs. We can solve this equation with Sommerfeld boundary conditions to obtain the radiation in the far-field region\revoyvind{, at a distance $r$ from the source,}

\begin{equation}
    A^i \stackrel{r\gg R}{\simeq}
    \frac{1}{\sqrt{4\pi} r}\int \diff ^3 x'\, j^i(t'_r,x'),
\end{equation}

\noindent
which, at first order in velocities, can be expressed as

\begin{equation}
 A^i  \stackrel{v\ll1}{\simeq} \frac{1}{\sqrt{4\pi} r} \left(S^i
    + n_j \dot S^{i,j}\right),\label{multipole}
\end{equation}

\noindent
with the moments defined as

\begin{equation}
S^{i,j\dots k} \equiv \int \diff^3 x' \,j^i x^j\dots x^k,\quad\text{and}\quad
Q^{i\dots m} \equiv \int \diff^3 x'\, j^0 x^i \dots x^m.
\end{equation}

\noindent
Due to conservation of the current $\partial_\mu j^\mu = 0$, we have the relations
\begin{equation}
    \dot Q = 0,\quad
    \dot Q^i = S^i = \mu\dot R^i(\sigma_1 - \sigma_2).
\end{equation}

%\begin{equation}
%\begin{aligned}\label{multipole}
 %   &\square A^i = j^i \implies A^i \stackrel{r>>R}{\simeq}
  %  \frac{1}{\sqrt{4\pi} r}\int \diff ^3 x'\, j^i(t'_r,x') \\&
   % \stackrel{v<<1}{\simeq}\frac{1}{\sqrt{4\pi} r} \left(S^i
    %+ n_j \dot S^{i,j}\right),
    %\\&
    %S^{i,j\dots k} \equiv \int \diff^3 x' \,j^i x^j\dots x^k,\quad
    %Q^{i\dots m} \equiv \int \diff^3 x'\, j^0 x^i \dots x^m\\&
    %\dot Q = 0,\quad
    %\dot Q^i = S^i = \mu\dot R^i(\sigma_1 - \sigma_2),
%\end{aligned}
%\end{equation}
%where we when inverting the d'Alembertian have used Sommerfeld boundary conditions, and below used the fact that $j^\mu$ is a conserved current, meaning $\partial_\mu j^\mu = 0$. 
\noindent
The first relation, stating charge conservation, guarantees the absence of monopolar radiation and that the leading order is given by the dipolar contribution. Note that if the two components carry the same charge-to-mass ratio, $\sigma_1=\sigma_2$, then the electric dipole vanishes and we need to go to the next order, whose term decomposes into the charge quadrupole and current dipole that generate electric quadrupolar and magnetic dipolar radiation respectively as the leading order contributions. An extension of our analysis that includes the second order term of the expansion is shown in Appendix \ref{chargequadrupole}.

Equipped with the radiation field, we can obtain the emitted energy per unit of time and solid angle by inserting the solution in the energy-momentum tensor $t^{\mu\nu}$. In the frame where the radiation propagates along the $z$-axis, we find

%We insert the above into the Noether current for translation to find the energy attributed to the EM field (in the frame where the radiation is along the z-axis):
\begin{equation}
    \frac{\diff P}{\diff \Omega} = r^2\langle t^{0z}\rangle
    = r^2\left\langle\dot A_x^2 + \dot A_y^2\right\rangle = \frac{\mu^2}{4\pi} (\sigma_1 - \sigma_2)^2\left\langle \ddot R_x^2 + \ddot R_y^2\right\rangle,
    \label{eq:dP/dOmega}
\end{equation}

\noindent
where $P=- \diff E/\diff t$ is the emitted power and $\langle f(t)\rangle$ is the average over one period, $\frac{1}{T}\int_0^T \diff t f(t)$.

Now, following \cite{maggiore_gravitational_2007}, we realise that we can rotate the dipole \revoyvind{from being along the $z$-axis} into an arbitrary direction $\hat{n}$ given by

\begin{equation}
   \hat{n} = \hat{\mathcal{R}}\hat{e}_z =  (\sin\theta\sin\phi,\sin\theta\cos\phi,\cos\theta)^T,
\end{equation}

\noindent
leaving all direction dependence to the rotation matrix. The general-direction dipole becomes

\begin{equation}
\begin{aligned}
    & \vec{Q}'(\hat{n}) = \hat{\mathcal{R}}^T \vec{Q}(\hat{e}_z) 
    =  \begin{pmatrix}
    Q_1 \cos\phi - Q_2 \sin\phi\\
    -Q_3 \sin\theta + \cos\theta( Q_1 \sin\phi + Q_2 \cos\phi) \\
    Q_3 \cos\theta + \sin\theta (
    Q_1 \sin\phi + Q_2 \cos\phi)
\end{pmatrix},
\end{aligned}
\end{equation}

\noindent
where we have used the rotation matrix given by

\begin{equation}
    \hat{\mathcal{R}} = \begin{pmatrix}
    \cos \phi & \sin \phi \cos \theta & \sin\phi \sin\theta  \\
    -\sin \phi & \cos \phi \cos \theta & \cos\phi \sin\theta \\
    0 & -\sin \theta & \cos \theta \end{pmatrix}.
\end{equation}

\noindent
We can then straightforwardly compute the emitted power by plugging these expressions into \eqref{eq:dP/dOmega} to obtain:

\begin{equation}
P = \frac{1}{3} \omega_0 \alpha^2\beta^3(2+e^2), \label{EMpower}
\end{equation}

\noindent
where we have defined the parameters

\begin{equation}
\begin{aligned}
    \alpha = a(1-e^2)\mu(\sigma_1 - \sigma_2),\quad
    \beta = (1-e^2)^{-3/2}\omega_0,\quad
    \omega_0 = \sqrt{\frac{\Tilde{G}M}{a^3}},
\end{aligned}
\end{equation}

\noindent
where the latter is the generalised Kepler's law that includes the electric potential effect through the effective Newton's constant $\Tilde{G}$.

Likewise, though a bit more involved, one can find the rate of angular momentum emission. We defer the details to Appendix \ref{AngularMomentum} and quote the final result here:

\begin{equation}
    \dot J = \frac{2}{3}\alpha^2\beta^2\omega_0.
    \label{angmompower}
\end{equation}

\noindent
Both of the above results agree with the expressions obtained in \cite{liu_merger_2020}\footnote{We thank the authors of \cite{liu_merger_2020} for helping us clarifying a disagreement  caused by a missing factor in their results. After this has been corrected we find perfect agreement.}, and with \cite{ld_landau__em_lifshitz_classical_nodate} for the particular case $G=e=0$.

We can likewise find the energy and angular momentum emission owed to gravitational waves, but refer to \cite{maggiore_gravitational_2007} for details. The power for a circular orbit, which we will use in \ref{linearised}, is:

\begin{equation}
    \label{gwpower}
    P_{GW} = \frac{32 G}{5}\mu^2 R^4 \omega^6,
\end{equation}

\noindent
where we have restored the gravitational constant, $G$, only here for being explicit about not promoting it to the effective gravitational constant $\tilde G$ used everywhere else in this work. $G$ is fixed by the requirement that the Schwarzschild metric should give an effective potential in the Newtonian limit equal to the Newtonian gravitational potential. The effective $G$'s rather stem from the dynamics of the orbit, from using our generalised Kepler's law in going back and forth between angular frequency and radius. In other words, in our final expressions for the total emitted gravitational radiation we will have two sources for the gravitational constant, namely: the coupling of the radiative gravitons to matter, that will contribute factors $G$, and the dynamics of the Keplerian orbits which, as we have seen, are determined by the effective gravitational constant $\tilde{G}$. Thus, by keeping track of the factors arising from the radiative and the non-radiative sectors we can easily identify the factors of $G$ and $\tilde{G}$ in the corresponding expressions.

\subsection{Evolution of Orbital Parameters}
We assume in the following that the emission of energy and angular momentum is dominated by the charge dipole.
Now that we know the emission rates, \eqref{EMpower} and \eqref{angmompower}, we can couple them to the time-differentiated equations for the energy and angular momentum of a Keplerian orbit, giving:

\begin{align}&
    \dot a(1-e^2) - 2a e \dot e =
    -\frac{B}{a^2\sqrt{1 - e^2}},\\&
    \dot a = - B \frac{1+ e^2/2}{a^2(1-e^2)^{5/2}},
\end{align}

\noindent
where we have defined

\begin{equation}
     B = \frac{4\mu \tilde G M (\sigma_1 - \sigma_2)^2}{3} .
\end{equation}

\noindent
We can solve the above equations analytically, following again \cite{maggiore_gravitational_2007}, to obtain

\begin{align}
    & \frac{\diff e}{\diff a} = \frac{3 e}{2 a}\frac{1-e^2}{2+e^2},\\
   & \frac{a}{a_0} = \frac{g(e)}{g(e_0)},
\end{align}

\noindent
where we have defined

\begin{equation}
    g(e) = \frac{e^{4/3}}{1-e^2},\quad\text{and}\quad
    \tau_{e=0} = \frac{a_0^3}{3 B},
\end{equation}

\noindent
with $\tau$ the time until collapse and collapse being defined as happening when $a=0$.

\subsection{Orbit Circularisation}
\label{circularisation}
\begin{figure}
\centering
\begin{subfigure}[t]{.49\linewidth}
    \includegraphics[width=\linewidth]{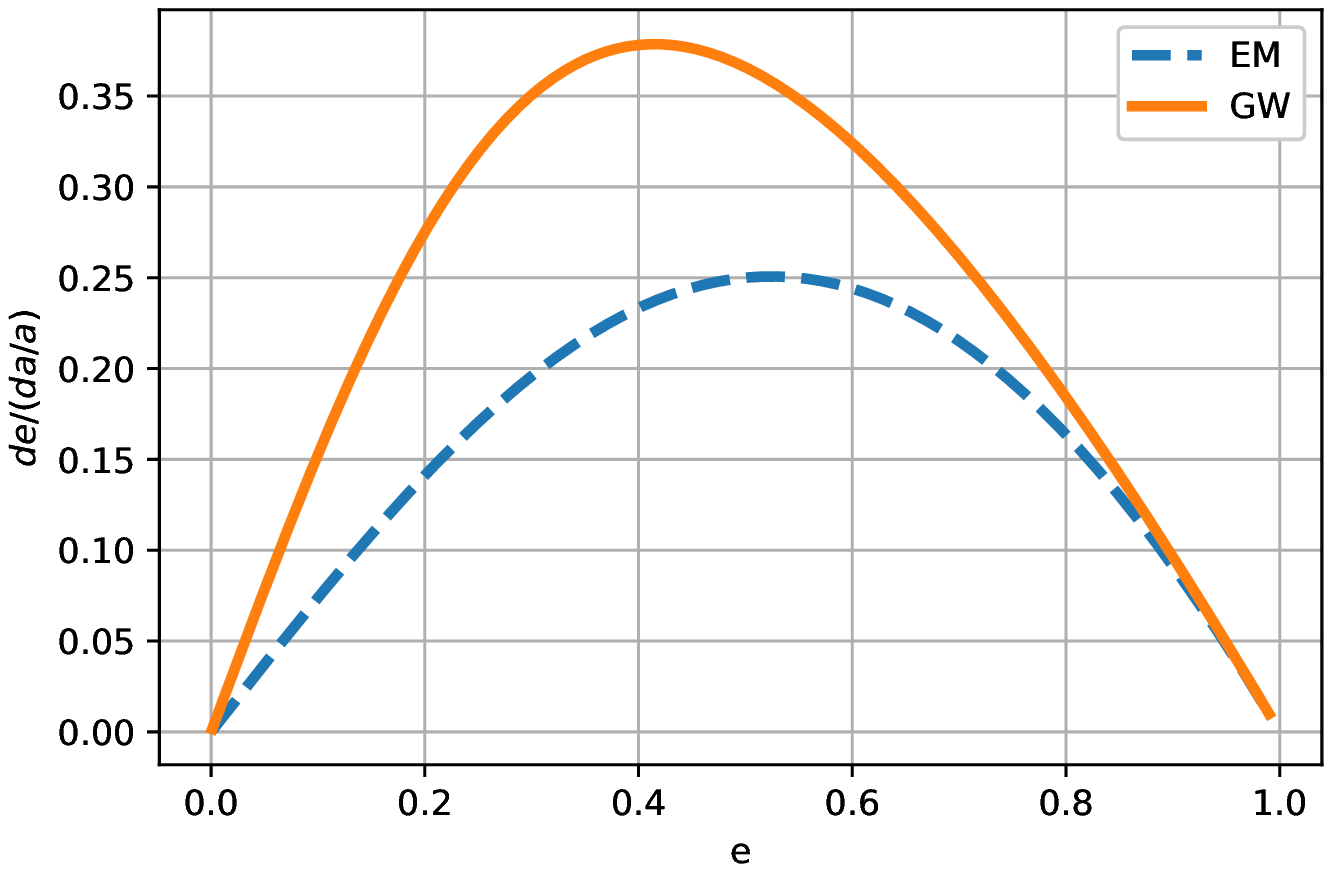}
\end{subfigure}
\begin{subfigure}[t]{0.49\linewidth}
    \includegraphics[width=\textwidth]{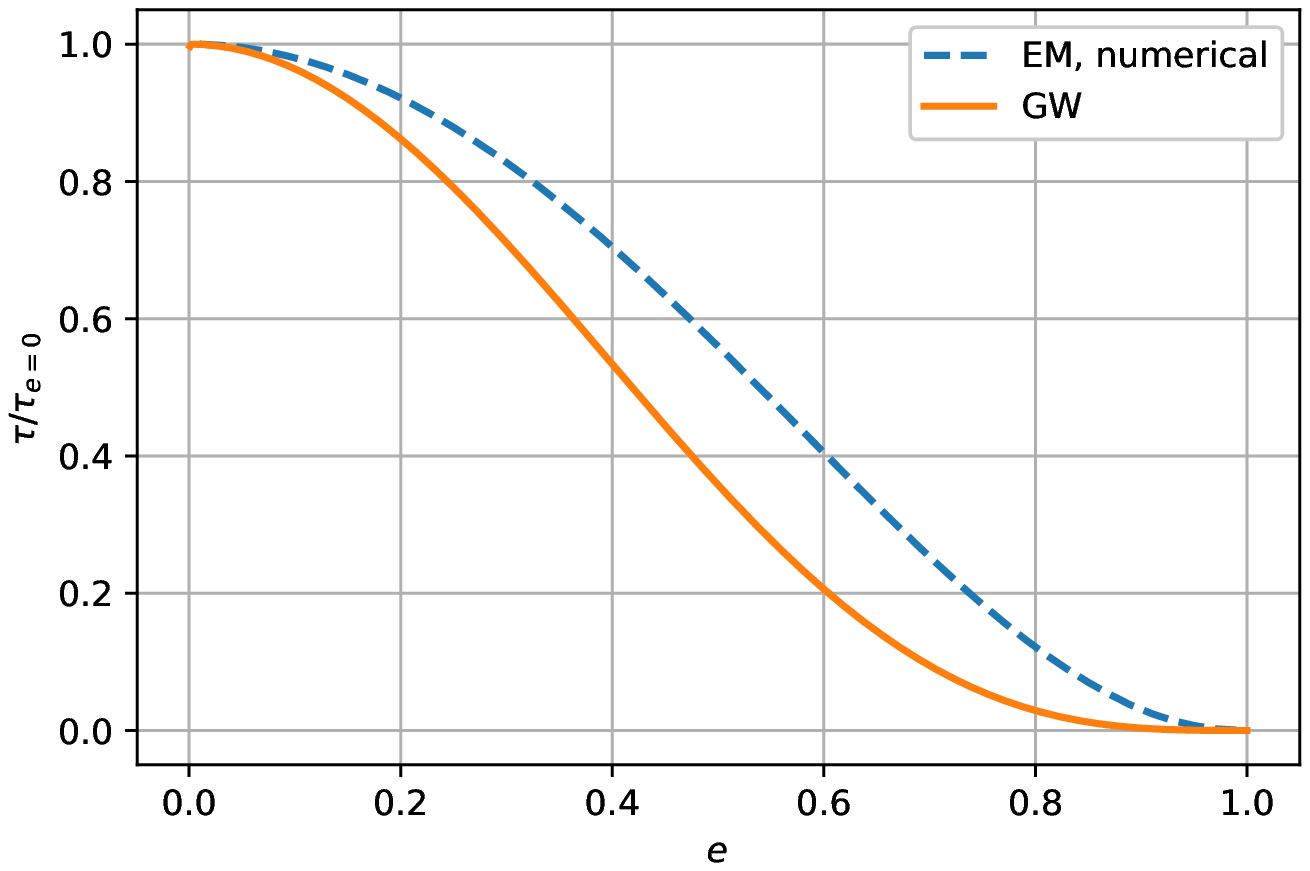}
\end{subfigure}
\begin{subfigure}[t]{.8\linewidth}
    \includegraphics[width=\linewidth]{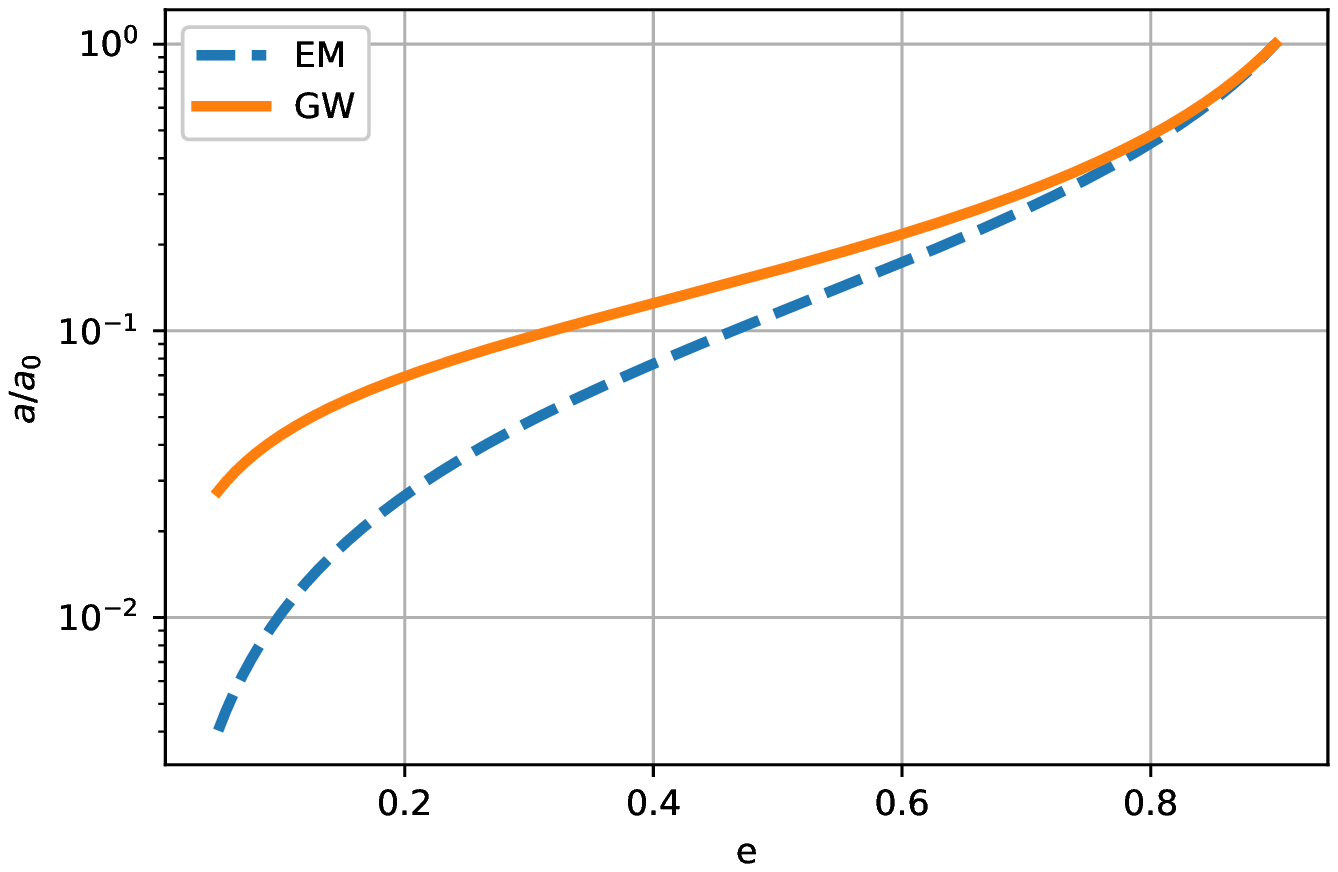}
\end{subfigure}
    \caption{Upper left: Change in eccentricity from a small variation in relative size of semi-major axis. Upper right: The time until collapse $\tau$. Bottom: Ratio of current to initial semi-major axis, with initial eccentricity $e_0=0.9$. All shown as functions of eccentricity, for both electromagnetically- and gravitationally dominated evolution of orbital parameters.}
    \label{Fig:Circularisation}
\end{figure}

We will now compare the results obtained in the previous section for the case of charged BHs to the corresponding results for the usual case of GW dominated emission. The equations for the latter can be found in \cite{maggiore_gravitational_2007} so we will not repeat them here. We show the results in Figure \ref{Fig:Circularisation}. We see that we can expect an initially (at large separation) eccentric black hole binary to be circularised by the time it enters LIGO/Virgo's sensitivity range when the GW channel dominates the energy and angular momentum emission, circularising within a reduction of about 1.5 orders of magnitude of the semi-major axis. For EM dominated emission, the circularisation is weaker, and requires a reduction of about 2.5 orders of magnitude.

In the case where the charge-to-mass ratios are equal so that the dipole is suppressed and the charge quadrupole is the leading order term, we found in Appendix \ref{chargequadrupole} that the emission is similar to the GW quadrupole emission up to a constant factor, equivalent to a rescaling of the chirp mass, otherwise given by $\mathcal{M} = \mu^{3/5} M^{2/5}$, and so the circularisation should be equally efficient as there is no dependence on the components' parameters.

\section{Linearised Case Parameter Estimation}
Now that we have justified our simplifying restriction of only considering circular orbits in our subsequent analysis, we turn to model the waveform and look at what sorts of errors on the chirp mass of the source an observer assuming pure Schwarzschild sources would get.

We note that templates matched onto, resulting from modelling of other effects, might be degenerate in parameter estimation with a charged template, and so a bias found in our consideration of only matching onto variations of initial phase and chirp masses might manifest differently if also including eccentricities, spin, dark matter halo dephasing \cite{macedo_into_2013}, sky- and polarisation angles, etc. We will in the following assume that there are no biases in the other parameters, focusing on the chirp mass estimation, under which assumption a sky- and polarisation angle averaged signal-to-noise ratio (SNR) consideration gives \revoyvind{the same} bias \revoyvind{as in our consideration} of a pure plus polarisation \revoyvind{hitting the detector straight on}. This can be seen using the same type of averaging as in chapter 7 of \cite{maggiore_gravitational_2007} when considering the sight distance or the stochastic background.

 Because of considering plus-polarised radiation hitting the detectors, the detector tensor when contracted with the perturbation tensor only results in an overall constant. The interesting part of the signal over which we will do our matching is then:
 
\begin{equation}\label{template}
    h = \omega^{2/3}\sin\Phi,
\end{equation}

\noindent
where $\omega$ is the orbital frequency and $\Phi$ is the phase, and both are functions of the binary parameters and time and will be found below.

We will in the following consider the small charge-to-mass ratio departure from the neutral case, linearising in a convenient charge parameter.

\subsection{Linearised Orbital Frequency and Phase}
\label{linearised}

We follow the same procedure as above, except that now we need only consider energy conservation to determine the orbital parameters as $e = 0$. When the charges (or their charge-to-mass ratio difference) are (is) small, both the GW quadrupole and the EM dipole emission can be important. In that case we need to couple the total power $P = P_{EM} + P_{GW}$ to the time differentiated energy equation. When doing this we find that

\begin{align}&\label{refthisequation}
    A_R \dot R = -1/R^3 - \epsilon_R/R^2,
\end{align}

\noindent
where we have defined the quantities:

\begin{equation}
    A_R = \frac{5}{64\mu(\tilde G M)^2 \Xi},\quad
    \epsilon_R = \frac{5(\Delta\sigma)^2}{48 \tilde G M \Xi},\quad
    \Xi = 1 + \mu^2(\sigma_1/m_1 + \sigma_2/m_2)^2/4,
\end{equation}

\noindent
where $\Xi$ comes from including the charge quadrupole emission. $\Xi$ is only a complete description of charge quadrupole order when the charge to mass ratio difference suppresses the dipole relative to the quadrupole, otherwise there would be a correction to the dipole from \spelloyvind{e.g.} the Biot-Savart interaction, and we should here restrict ourselves to dipole order, setting $\Xi=1$. The details can be found in Appendix \ref{chargequadrupole}.

The solution to \revoyvind{equation \eqref{refthisequation}} can be written as:

\begin{equation}
\begin{aligned}&
    \frac{t}{A_R} = f(R_0) - f(R),\quad 
    \tau(R) = A_R f(R),
\end{aligned}
\end{equation}

\noindent
where \revoyvind{$R=R_0$ at $t=0$, $\tau$ is the time until coalescence, and}

\begin{equation}
    f(R) = - \frac{\ln (1+\epsilon_R R)}{\epsilon_R^4} + \frac{R}{\epsilon_R^3} - \frac{R^2}{2\epsilon_R^2} + \frac{R^3}{3\epsilon_R}
    \stackrel{\epsilon_R R \ll 1}{\simeq}
    R^4/4 - R^5 \epsilon_R/5.
\end{equation}

\noindent
Linearising in $\epsilon_R R = \epsilon \omega^{-2/3}$, we eventually find ($u=\tau/\tau_0$):

\begin{align}&
    \frac{R}{R_0} = u^{1/4}\left[
    1 - \frac{R_0\epsilon_R}{5}\Big(1-u^{1/4}\Big)\right],\quad
    R_0 = \left(\frac{4\tau_0}{A_R}\right)^{1/4}\left[1+\frac{\epsilon_R}{5}\left(\frac{4\tau_0}{A_R}\right)^{1/4}\right],\\&
    \frac{\omega}{\omega_0} = 
    u^{-3/8}\left[1 + \frac{3}{10}
    \epsilon\omega_0^{-2/3}\left(1-u^{1/4}\right)
    \right],\quad
    \omega_0 = \left(
    \frac{3 A}{8\tau_0}\right)^{3/8}
    \left[1 - \frac{3}{10}\epsilon
    \left(\frac{8\tau_0}{3 A}\right)^{1/4}\right], \label{linfrequency}
\end{align}

\noindent
\revoyvind{where we have made the useful definitions}

\begin{equation}
        A = \frac{5}{96\mathcal{M}_*^{5/3}},\quad
    \epsilon = \frac{5\mu(\Delta\sigma)^2}{48\mathcal{M}_*^{5/3}},\quad
    \mathcal{M}_* = (\Xi \mu)^{3/5}(\tilde G M)^{2/5},
\end{equation}

\noindent
\revoyvind{
and $\mathcal{M}_*$ is the generalised chirp mass.}

We can now find the phase. Setting $\Phi_0 = 0$, we obtain

\begin{equation}
\begin{aligned}
    &\Phi(t) = \int_{t_0}^t \omega_{gw} \diff t' = 2 \int_{t_0}^t \omega\, \diff t' ,
\end{aligned}\label{genphase}
\end{equation}

\noindent
and so we \revoyvind{define}

\begin{equation}
     \tilde \Phi \revoyvind{=} \frac{5 \Phi}{16 \tau_0}\left(\frac{8\tau_0}{3 A}\right)^{3/8} =
    1 - u^{5/8} -
    q\left(1-  u^{7/8}\right), \label{linphas}
\end{equation}

\noindent
with 

\begin{equation}
q = \frac{3\epsilon}{14}\left(\frac{8\tau_0}{3 A}\right)^{1/4} 
     = \frac{5}{56}\mu(\Delta\sigma)^2\left(\frac{\tau_0}{5\mathcal{M}_*^5}\right)^{1/4} 
     \label{qquant} .
\end{equation}

\noindent
We can then construct our templates by inserting \eqref{linfrequency} and \eqref{linphas} into \eqref{template}.

For the Fourier phase of the plus polarisation, using equation (4.367) in \cite{maggiore_gravitational_2007}, we find

\begin{equation}
    \Psi_+ =
    2 \omega (\revoyvind{\tau_0} + r) -\pi/4 - \Phi_0
    +\frac{3}{4}\left(8\mathcal{M}_* \omega\right)^{-5/3}
    - \frac{5\mu (\Delta \sigma)^2}{28 \mathcal{M}_*}
    \left(8\mathcal{M}_* \omega\right)^{-7/3},\label{Fourierphase}
\end{equation}

\noindent
which correctly reduces to (4.37) in \cite{maggiore_gravitational_2007} for the uncharged case and agrees with equation (3.8) in \cite{cardoso_black_2016}\footnote{We thank the authors of \cite{cardoso_black_2016} for helping us to clarify some discrepancies we found with their results due to some typos and missing factors in their analysis.}.

\subsection{A Least Squares Approach}

We compare the template constructed above for some small $(\Delta\sigma)^2$ to another one of the above with $\Delta\sigma=0$ and a rescaled chirp mass $\mathcal{M}_*\rightarrow a\mathcal{M}_*$, holding $\tau_0$ fixed. We then determine the rescaling of the generalised chirp mass, $a$, by finding the least squares of their phase difference over all $u$. For simplicity we only match their phases, as this is the most important effect. The following will then serve as an approximation and setting of expectations of what we will find later when considering the full matched filtering template selection.

We thus have 2 different transformations of the phase that we wish to match to each other

\begin{align}&
    \delta_w \tilde \Phi = - (1-w)\cdot \left(1 - u^{5/8}\right),\quad
    \delta_q \tilde \Phi = - q\cdot \left(1 - u^{7/8}\right),\\&
    \delta_w\tilde\Phi - \delta_q\tilde\Phi
    \stackrel{\text{match}}{=} - \tilde \Phi_0,\quad
    w = 1/\revoyvind{a}^{5/8},\quad t\revoyvind{(c)}=(1-w)/q,\quad s\revoyvind{(c)}=\tilde\Phi_0/q .
\end{align}

\noindent
Putting up the least squares equation

\begin{equation}
    \begin{pmatrix} \frac{\partial}{\partial t} \\
    \frac{\partial}{\partial s}\end{pmatrix}
    \int_c^1 \diff u \left[t(1-u^{5/8}) - 1 + u^{7/8} - s
    \right]^2 = 0 ,
\end{equation}

\noindent
we find the solutions

\begin{align}&\label{tfunc}
    t(c) = \frac{\frac{8}{15} c^{15/8} - \frac{2}{5}c^{5/2} - \frac{2}{15} +
    \frac{8(1-c^{15/8})}{15(1-c)}(\frac{5}{13}-c +\frac{8}{13}c^{13/8})
    }{
    \frac{25}{117} - c+ \frac{16}{13}c^{13/8}-\frac{4}{9}c^{9/4} - \frac{(\frac{5}{13} + \frac{8}{13}c^{13/8} - c)^2}{1-c}
    },\\&\label{sfunc}
    s(c) = \frac{t(c) ( \frac{5}{13} +\frac{8}{13}c^{13/8} -c ) + \frac{8}{15}(1-c^{15/8})}{1-c} - 1,
\end{align}

\noindent
where $c$ is the cutoff, motivated by the fact that our weak-field low-velocity approximations will fall apart for small $u$, invalidating our analysis. We therefore only perform the matching for $u\in(c,1)$.

The above gives for the parameter estimation:

\begin{align}\label{genphasebias}
    &\frac{\mathcal{M}_*^{\text{new}}}{\mathcal{M}_*^{\text{true}}}
    \simeq 1+ \frac{8 q}{5}  t(c),\\&
    \frac{\mathcal{M}^{\text{new}}}{\mathcal{M}^{\text{true}}}
    \simeq \tilde G^{2/5}\Xi^{3/5}  \left( 1+ \frac{8 q}{5}  t(c) \right)
    \label{bias1},\\&
    \Phi_0 = \frac{16\tau_0}{5}\omega_0 q s(c)\label{bias2},
\end{align}

\noindent
where $q$ is defined in \eqref{qquant}. These results clearly signal that the obtained parameters, i.e. the chirp and total masses, will be biased with respect to the true values of the emitting source. This puts forward the risk of a mismatch between the true mass of the source and the inferred one when we try to fit a charged binary system to a binary system of Schwarzschild BHs. In the next section we will confirm this result with a more precise and robust numerical approach.

\subsection{Matched Filtering and Numerics}
\label{numericsInText}
\begin{figure}
    \centering
    \includegraphics[width=0.8\textwidth]{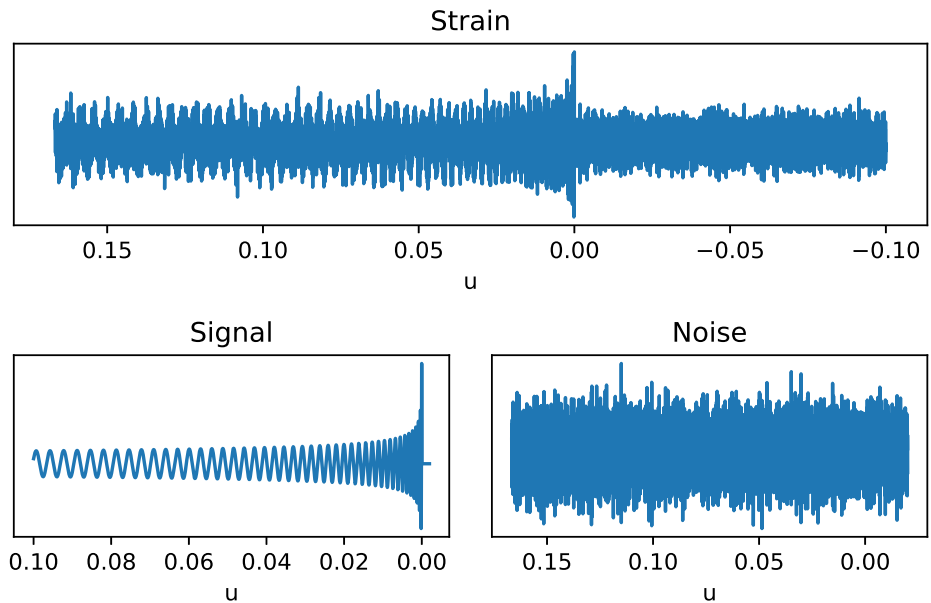}
    \caption{The strain, signal and noise as functions of $u$. Relative strain/noise amplitude, cutoff \oyvind{and coalescence time} chosen here for illustrative purposes.}
    \label{fig:strain}
\end{figure}

We will start with the template \eqref{template}, normalised to unity at some small time before the merger and inserted into it our linearised expressions for frequency \eqref{linfrequency} and phase \eqref{linphas}, and populate it with stationary, \revoyvind{aLIGO noise taken from the design sensitivity interpolation of \cite{arun_parameter_2005} } -- the result is our strain, seen in figure \ref{fig:strain}. We then construct a grid of charge-neutral templates, varying over different generalised chirp masses and initial phase constants, to each point in which we attach an SNR-value (signal-to-noise ratio) that quantifies how good the match is. The selected template is the one with the highest SNR, and its parameters will be what is deduced for the emitting source.

For our cutoff, $c$, we need to have simultaneously

\begin{align}
    v \ll 1,\quad R \gg 3 R_s,\quad T \dot R/ R \ll 1,
\end{align}

\noindent
where \revoyvind{$v$ is the velocity of the effective 1-body motion}, $T$ is the period, $R_s=2M$ is the summed Schwarzschild radii, used here to approximate the summed outer Reissner-Nordström radii for the weakly charged black holes and $3 R_s$ is the corresponding innermost circular orbit. These conditions correspond to the assumptions of low-velocity, weak-field or circular orbit, and quasistatic inspiral respectively. We find that the second condition is the strictest one, giving

\begin{equation}
    u > \oyvind{25.3} \frac{\mathcal{M}}{\eta^{8/5}\tau_0},
\end{equation}

\noindent
where $\eta=\mu/M$ is the symmetric mass ratio, which at largest, for equal component masses, is $1/4$, giving \oyvind{ $u > 0.17$} when we pick $\mathcal{M}=28.6 M_\odot$ and $\tau_0 = 0.2$s which were the case for GW150914. For $m_1/m_2=10$, we find $u > 1$ which \oyvind{renders our method useless}. We \spelloyvind{therefore} pick \oyvind{$u=0.2$} \spelloyvind{there} and note that the treatment becomes invalid for asymmetric masses. For our other event considerations, see table \ref{tab:parameterchanges}. We have some room to add asymmetric masses, but keep the symmetric $\eta$ limit for easier comparison among them. \revoyvind{We do not find a large sensitivity of our results to the cutoff. When we put $\mu\Delta\sigma=0.6$ and consider the Figure \ref{fig:chirp} scenario, we see that cutoffs between 0 and 0.2 give an at most $\sim 2 \%$ difference on the SNR fraction and a smaller than $\sim 0.5 \%$ difference on the chirp mass fraction, meaning that for our choice of $u=0.02$ in figure \ref{fig:chirp}, we seem to suffer negligible effect from our rough cutoff. In the two other scenarios, we get a rather large variations in the SNR ratios and find for GW151226 and GW150914 respectively $20 \%$ and $30 \%$ difference at their respective cutoffs compared to $c=0$. The same numbers for the chirp ratios remain small though, at $1 \%$ and $3 \%$ respectively.} 

\oyvind{
We also require the linearisation parameter $\epsilon \omega^{-2/3}$ to at largest be equal to $0.1$. This amounts to, after some algebra,

\begin{align}
    \left[ \mu \left( \Delta\sigma\right)^2\right]_{\text{max}} = \frac{ 6 \mathcal{M}^{5/4}}{25} \left( \frac{15}{2\tau}\right)^{1/4}, \label{chargelimit}
\end{align}

\noindent
which \spelloyvind{suggests} that smaller observing times and larger masses means that our linearisation scheme is valid for a larger domain of charge to mass ratios.}

We note that a more thorough analysis might seek to vary both the initial time to coalescence $\tau_0$ and the cutoff $c$ as functions of the signal's parameters due to small-charge corrections to frequency (the detectors having a lower frequency threshold) and circular orbit/weak-field limits. We believe that this would constitute part of a more thorough analysis where the complete waveform is considered.

Our code for finding the SNR of a given signal and template is built around the example code provided at the LIGO open science center \cite{vallisneri_ligo_2015}. \label{Appendix:matchedfilteringnumerics} \revoyvind{
We provide more information about our implementation of the matching procedure in appendix \ref{Appendix:matchedfilteringnumerics} and provide general quantities used in the code in table \ref{table:quantitiesnumerics}, and simulations specific ones in tables \ref{tab:parameterchanges} and \ref{table:quantities_simulationsspecific}. } 
\subsection{Results}
\label{results}

\begin{figure}
    \centering
    \includegraphics[width=0.8\textwidth]{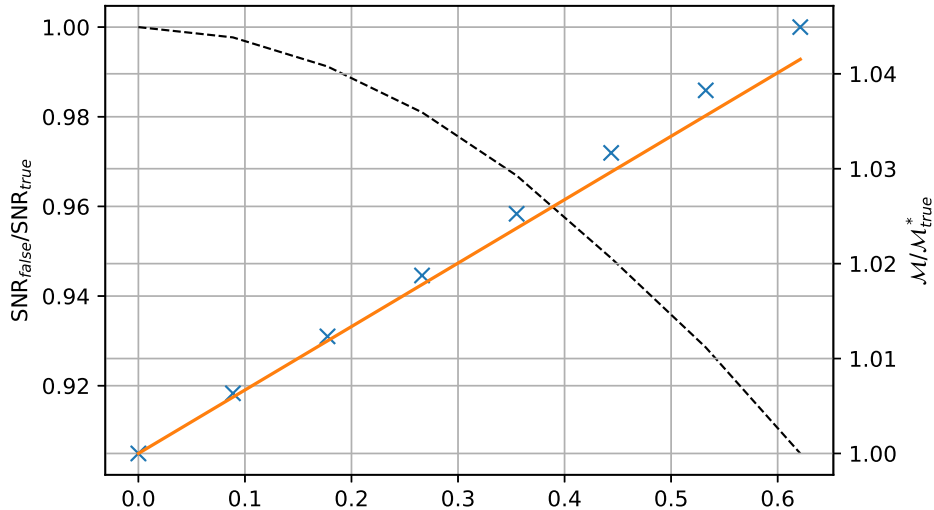}
    \caption{The overestimated generalised chirp mass (right axis) as a function of the $\mu(\Delta\sigma)^2/M_\odot$ parameter of the source. Result from analytically applying method of least squares to the phases in solid line, and results from matched filtering marked by crosses. The SNR for the false match relative to what would have been the case if the true template was contained in the template bank (left axis) in dashed line. Chosen here is a true generalised chirp mass of 30 solar masses and an initial time until coalescence of \oyvind{20 seconds, with a cutoff of $c=0.002$} and the noise parameter $N=\sqrt{S_0}=1$ (see code). Results are averaged\oyvind{ over 20 realisations.}}
    \label{fig:chirp}
\end{figure}

The result of the matched filtering analysis, for the special case of $\mathcal{M}=30 M_\odot$ and \oyvind{ $\tau_0 = 20$s }, can be seen in Figure \ref{fig:chirp}.\oyvind{ Although the typical observation time of such large-mass mergers is significantly smaller, see \cite{ligo_scientific_collaboration_and_virgo_collaboration_gwtc-1_2019}, we pick it here motivated by future observations with increased sensitivity and lower frequency thresholds, like we will expect from the LISA mission. This choice also gives a clearer chirp mass bias relation than the smaller observation times.} \revoyvind{We average over 20 simulations, but find basically no spread in the estimations, asummedly owing to the long signal. We did not find any significant spread in the result until we use $N\sim$ amplitude of noise/signal in time-series $\sim 10^{8}$, and can therefore safely assume convergence in the figure, where $N=1$. For the other scenarios we found some significant spread already at $N\sim 1$ and growing significantly for $N\sim 10$ and averaged therefore over a greater amount of realisations.}

We note that the slope of the least squares result is somewhat off, and attribute this to the fact that we only considered the phase in that analysis. \oyvind{We provide some numbers for different parameter choices in table \ref{tab:parameterchanges}. }

\begin{table}
    \centering
    \begin{tabular}{|c|c|c|c|} \hline
    Parameters & $\mathcal{M}$ ($M_\odot$) & $\tau_0$ (s) & c \\ \hline
    GW151226 & 8.9 & 1.7 & 0.01 \\
    GW150914 & 28.6 & 0.2 & 0.2 \\
    Figure \ref{fig:chirp} & 30 & 20 & 0.002 \\ \hline
     Results & $ (\mathcal{M}_*/\mathcal{M}_*^{\text{true}} )_{\text{max} }$
     & $(\text{SNR}/\text{SNR}^{\text{true}})_{\text{min / max}}$ & $\mu \left(\Delta\sigma\right)^2_{\text{max}} \,(M_\odot)$\\
     \hline
    GW151226 & 1.04 & 0.999 / 1 & 0.25 \\
    GW150914 & 1.047 & 0.8 / 1.3  & 1.78 \\
    Figure \ref{fig:chirp} & 1.045 & 0.905 / 1 & 0.62 \\ \hline
    \end{tabular}
    \caption{\oyvind{Results for different choices of parameters. $\mathcal{M}_*$ is the generalised chirp mass, $\tau_0$ the coalescence time, $c$ the cutoff and SNR the signal-to-noise ratio. True is used about the quantities that would have been found if the charged templates had been contained in the matching pipeline. The results are found from an average of 1200, 6500 and 20 runs respectively (top-down).}}
    \label{tab:parameterchanges}
\end{table}

For a generalised chirp mass of 30 solar masses, in the worst considered case of \oyvind{ $\mu(\Delta\sigma)^2/M_\odot=0.62$}, we would then infer a generalised chirp mass of \oyvind{31.35} solar masses. If we assume the component masses to be equal, and only one of the components charged, so that $\tilde G = 1$, then the reduced mass would be \oyvind{18.01} solar masses. \spelloyvind{T}he charge to mass ratio of the charged body \spelloyvind{would be} $\sigma_1 \sim 0.19 \sim 1.6 \times 10^{-11} $C/kg and total charge $q_1 \sim 10^{21}$C. This would then cause an overestimation of the component mass of 34.46 solar masses to \oyvind{36.01} solar masses.

We should however note that $\tilde G = 1-\sigma_1\sigma_2$, and so if both bodies are charged positively, we might potentially get an underestimation of the mass when we are interpreting the generalised chirp mass as the actual chirp mass, because $\mathcal{M}_* = \tilde G^{2/5}\Xi^{3/5}\mathcal{M} $. If they both are charged equally, the dipole would disappear, but this bias would remain in the GW quadrupole as the modification is due to the generalised Kepler's law and charge quadrupole radiation. Then, with say $\sigma_1=\sigma_2=0.19$, we would get $\frac{\mathcal{M}^{\text{new}}}{\mathcal{M}^{\text{true}}}=0.99$, which is less significant than what was found above\spelloyvind{, but} will work to reduce the chirp mass bias in the equal-sign-charge, non-vanishing dipole case.

Another interesting detail is that according to equation \eqref{genphasebias}, we would expect $\mathcal{M}^{\text{true}}_*/\mathcal{M}^{\text{false}}_* -1 \propto \tau_0^{1/4}$, and so we should expect a greater bias the longer before the merger the strain enters our detectors' sensitivity. 
We also see, again according to equation \eqref{genphasebias}, that smaller true chirp mass mergers will have greater biases. An interesting effect of the dependence of $\tau_0$ on the biased chirp mass is that there will be an additional relation between the bias factor and the true mass related to the experimental setup. This relation arises because the LIGO/Virgo interferometers have a finite frequency band with a lower limit that determines when a given signal will enter the detector for a given mass and this lower limit will also determine the value of $\tau_0$. \oyvind{ We can use the equation for $\omega_0$ \eqref{linfrequency} to zeroth order in charge, solve for $\tau_0$ and insert it back into \eqref{bias1} to find how the dipole induced bias depends upon the detector threshold. We find

\begin{align}
    \frac{\mathcal{M}_*^{\text{new}}}{\mathcal{M}_*^{\text{true}}} = 1 + \frac{t(c)}{28}\left(\Delta\sigma\right)^2 \left(M \omega_{\text{detector}}\right)^{-2/3},
\end{align}

\noindent
where $\omega_{\text{detector}}$ is the lower frequency sensitivity of the detector (assumed that it is the frequency range of the detector, and not the amplitude of the signal that determines when the signal first becomes detectable). We see then that the smaller the lower frequency, the larger the bias, while smaller masses also work to increase the bias. On the other hand, looking at equation \eqref{chargelimit}, we see that the charges where our linearisation scheme is valid decrease with these same parameters, which we experience in our choice of parameters to conspire to keep the maximum bias always around $\sim 5 \%$.
}

We do not find a loss of SNR for scenarios considered here of much more than a few percent, agreeing with \cite{khalil_hairy_2018}, which points to most such weakly charged mergers actually being detected alongside uncharged mergers, though they might not be considered significant and their chirp masses would be estimated with a bias. \oyvind{For the parameter choice according to GW150914 however, we find both a big loss and big gain of SNR for different true values of our $\mu\left(\Delta\sigma\right)^2$ parameter.  We attribute this to the shortness of the signal, the big cutoff and the resulting small signal of the matching. \revoyvind{This notion is supported by the fact that the SNR ratio varied significantly as we varied the cutoff for \spelloyvind{the two short-signalled} scenarios.}}

\section{Hyperbolic Encounters}
\label{hyperbolic}

After studying the case of a bound system, we now would like to extend our previous analysis to a treatment of hyperbolic encounters similarly to what is done for the uncharged case in \cite{garcia-bellido_gravitational_2017,garcia-bellido_gravitational_2018}.

\subsection{Adiabatic Approximation}
An immediate complication with respect to the bound case is the fact that there is no analogue to the quasistatic orbit approximation, where we average the emission over an orbital period, considering the orbital parameters to be changing at a larger timescale. For the hyperbolic encounter there is no periodicity in the motion to average over, and so we are faced with the Weinberg-Witten theorem \cite{weinberg_limits_1980} stating that  one cannot build a Lorentz-covariant and gauge invariant stress-energy tensor for a massless particles with spin larger than 1, such as the graviton.

There then seemingly is an irremovable gauge-ambiguity to asserting back-reaction on the orbit from graviton emission, though we can still consider electromagnetically dominated emission without problems. The authors of \cite{garcia-bellido_gravitational_2017} tackled the ambiguity by treating static orbits, assuming the orbital parameters to change on a timescale larger than that of the encounter. There will however still be an ambiguity to the instantaneously emitted power and the frequencies of the radiation, but we consider this irrelevant since our real observable is the proper length difference in the arms of the detectors in the detector frame, which through Fermi normal coordinates is expressible with the Riemann tensor which is gauge invariant in the linearised theory \cite{maggiore_gravitational_2007}.

For the orbit, a first guess on the back-reaction could be to approximate the interaction as all happening at the periapsis, so that there is an initial and final hyperbolic trajectory differing in energy and angular momentum by what is found by integrating the emission over the entire initial static orbit. Another solution could be to perform window-averaging over several periods of the frequency of interest and assign the power to the centre of the window.

\subsection{Static Trajectories}
Putting aside the aforementioned considerations, we will now consider static trajectories without the averaging discussed above, as in \cite{garcia-bellido_gravitational_2017}. The generalisation to U(1) charged mergers only brings with it a modification to the relations between orbital parameters $(a,e)$ and the components' parameters. We find

\begin{equation}
    \begin{aligned}&
    e = \sqrt{1+\frac{b^2}{a^2}}
    =\sqrt{1 + \frac{b^2 v_0^4}{\tilde G^2 M^2}},\\&
    a = \frac{\tilde G M \mu}{2 E} = \frac{\tilde G M}{v_0^2},\\&
    R = \frac{a(e^2-1)}{1+e\cos(\psi-\psi_0)},
    \end{aligned}
\end{equation}

\noindent
where $b$ is the impact parameter or the semi-minor axis.

The remaining development in \cite{garcia-bellido_gravitational_2017} resulting in their equations (7-16) then follow straightforwardly, luckily being expressed in terms of $(a,e)$. There is however a small modification to the constraints put in their equations (4-6). Instead of requiring that their closest distance is larger than their summed Schwarzschild radii $r_{\text{min}}>\,_1R_s + \,_2 R_s \equiv R_s$, we need to impose that it is larger than the corresponding outer Reissner-Nordstr\"om radii $R_+$, and so we find

\begin{equation}
\begin{aligned}&
    r_{\text{min}} = a(e-1) = b \sqrt{\frac{e-1}{e+1}}>\,_1R_+ + \,_2 R_+ \\&\equiv \frac{R_s}{2} + m_1\sqrt{1 - \sigma_1^2}
    + m_2\sqrt{1 - \sigma_2^2}.
\end{aligned}
\end{equation}

\noindent
This may or may not be a stricter requirement than $v_{\text{max}}<1$, which leads to

\begin{align}&
    b > \frac{(e+1)^{3/2}}{(e-1)^{1/2}}\frac{\tilde G R_s}{2}.
\end{align}

\noindent
Both of these are for the minimally charged case approximated by the uncharged conditions of \cite{garcia-bellido_gravitational_2018}. The net effect of this whole consideration is that the five independent parameters become $(M,\mu,v_0,b,R)\rightarrow(\tilde G M, \mu, v_0, b, R)$. Note again that the $G$ appearing in Einstein's field equations is still just the normal gravitational constant, as discussed in section \ref{model}. Equations (56-60) of \cite{garcia-bellido_gravitational_2018} then follow with $M\rightarrow\tilde G M$ which is the same type of substitution we have done in our generalised chirp mass above, not taking into account back-reaction. There then is a degeneracy in estimation of $\tilde G$ and $M$ that potentially can be broken by a careful modelling of the back-reaction of the radiation on the orbit.

\section{Conclusions and Discussion}

In this work we have considered the emission of GWs by systems involving charged BHs whose charge is not the usual electromagnetic charge, but rather corresponds to some dark (hidden) charge. We have studied the cases of a bound system of two black holes and an unbound system corresponding to a hyperbolic encounter of two of such objects. 

In Section \ref{circularisation} we showed that an EM dipole dominated emission of a black hole binary in Keplerian orbits would cause the orbits to circularise by the time the components' separation has decreased a considerable fraction. We then verified that only considering circular orbits by the time the mergers enter into the LIGO/Virgo sensitivity might be a good approximation for EM radiating binaries as well as binaries with GW dominated emission. This would depend on their initial separation which in turn depends on the specific binary formation process considered.
In Section \ref{results} we saw that we could get a \oyvind{somewhat} significant \oyvind{($\sim 5 \%$)} overestimation for the chirp mass of a binary system when projecting a charged signal onto uncharged templates within our linearisation scheme, but it would require a somewhat high amount of charge of $\sigma \sim 0.1$, which seems to agree with \cite{zhu_inspirals_2018}. We saw that, according to our results, the weakly charged mergers would likely be detected alongside uncharged mergers \oyvind{(although this was unclear for very short signals)}. Finally, in Section \ref{hyperbolic}, we did not find the bias due to emission of U(1) radiation for the hyperbolic encounters, but did find a modified relation between the parameters of the static orbit and the components. We saw that, not considering back-reaction, the main difference is caused by the modified Kepler's law and hides away in the total mass parameter of the assumed uncharged encounter. \revoyvind{Our work then suggests that the inclusion of charge could be related to} \revoyvind{the} large estimated chirp masses, which \revoyvind{is one} of the main unexpected features about the detected binaries. \revoyvind{In this respect, similar effects can be expected for other exotic charges. The biases found here are however not big enough to do this alone, and we therefore would need the bias we are seeing to keep increasing with charge, beyond our linearisation regime, without reducing the SNR too much so that we do not compromise the detections.}

In a detour from our main focus of a hidden charge, we would like to comment now on the impact of our results for the mechanism suggested in \cite{wald_black_1974} through which a spinning black hole in a uniform external magnetic field would naturally acquire electric charge, which for a Kerr BH would give at most a charge of:

\begin{equation}
    \sigma = \frac{q}{m} \ll2 B_0 m = 1.7\times 10^{-20}\left(\frac{m}{M_\odot}\right)B_0(\text{Gauss}),
\end{equation}

\noindent
which we see is completely negligible for our purposes for typical galactic magnetic field strengths of order micro Gauss \cite{beck_galactic_2007}. To get a $\sigma=0.19$, for $m_1 = 34.46 M_\odot$, we would need a magnetic field of order $10^{17}$ Gauss, which not even living on top of a magnetar could account for, typical field strengths there being at most of order $10^{15}$ Gauss \cite{kaspi_magnetars_2017}. There could however, for all we know, be some other unknown mechanisms that would allow for more electric charge, for example if dark matter carries fractional electric charge, and if we think of the U(1) charge as a hidden charge then there are more possibilities; If the universe is not neutral in this other charge, for example, the BH might be charged from the onset. For a discussion on hidden and on fractional electric charge, see \cite{cardoso_black_2016}. For new constraints on electrically minicharged dark matter, see \cite{plestid_new_2020}.

The above has a very restricted scope owing to all of its simplifying assumptions. Concerning sources, it would be illuminating to see a similar analysis done for either or both of spinning black holes and without linearising in the charges. Regarding the false match parameter estimation it would be interesting to consider biases in other parameters like residual eccentricities, spin, sky- and polarisation angles and dark matter halo dynamical friction \cite{macedo_into_2013}.

As mentioned in section \ref{results}, based on equation \eqref{genphasebias}, the bias is expected to both grow with initial time until merger and shrink with the true, generalised chirp mass, and we therefore would expect neutron star mergers, that are better in both these regards, to make an interesting subject for further studies if they could acquire some hidden charge. This is speculative because an appropriate mechanism for the neutron stars to maintain a non-negligible charge should first be devised. \oyvind{Also, as we mentioned in the results section, equation \eqref{chargelimit} constrains our charge choices more for such scenarios, keeping our max bias nearly constant. We thus would require higher order corrections in order to consider the same charges that we do for the black holes.}

Other ideas for future work is to look at a similar type of analysis for non-Abelian charges or to model the back-reaction of both the (dark) photon and graviton radiation on the hyperbolic trajectories to see whether there is some exciting new phenomenology and if the mass degeneracy mentioned may be broken.

Considering Figure 4 in \cite{flanagan_measuring_1998}, we see that our analysis might give a correct notion of the bias introduced in the LIGO detectors for total masses below $20-30 M_\odot$, while for larger masses, first the merger- and then the ringdown phase dominate with their contribution to the SNR, while the inspiral becomes of decreasing importance. It would therefore be interesting to see what kind of mass bias we could get from the merger phase. The ringdown is covered in \cite{cardoso_black_2016}. \revoyvind{Going to these different regimes could help break the mass/charge degeneracy as we do not expect the presence of charge to be more or less absorbable into the other parameters in general, and even if they were, the biases in the different regimes would presumably differ. Doing checks such as in \cite{the_ligo_scientific_collaboration_and_the_virgo_collaboration_tests_2019} might therefore be a fruitful endeavour. Going to higher post-Newtonian orders in the inspiral might also help in this, as the different order terms assumedly also would absorb parameters differently, but whether the biases cancel or not ought to be checked explicitly, and our first order consideration works as a proof-of-concept that they might not.}\newpage

Finally, it would be interesting to see the construction and inclusion into the LIGO/Virgo template banks of charged components mergers, to estimate and put constraints on the charges and possibly remove parameter estimation biases -- this would however require a higher expansion in velocity for the inspiral, numerical simulations for the merger phase, and solutions of the quasi-normal modes of a Kerr-Newman black hole for the ringdown, all to a high precision. In view of our results we believe this is a pertinent endeavour to pursue in order to have an optimal and more appropriate exploitation of GWs data.
\newline

\section*{Acknowledgement}
We would like to thank Alicia Sintes, Jonah Kanner, Fernando Atrio, Daniel Heinesen, Alex Ziegenhorn, Espen Christiansen and Liu Lang for useful comments and/or discussions. We thank the authors of \cite{cardoso_black_2016} and \cite{liu_merger_2020} for their help in comparing our results with theirs. We also thank the authors of \cite{garcia-bellido_gravitational_2017} for discussions on hyperbolic encounters. ØC thanks the department of fundamental physics at the university of Salamanca for hosting him during this work.
We thank the Research Council of Norway for their support.  The simulations were performed on resources provided by 
UNINETT Sigma2 -- the National Infrastructure for High Performance Computing and 
Data Storage in Norway. JBJ acknowledges support from the  {\textit{ Atracci\'on del Talento Cient\'ifico en Salamanca}} programme and the MINECO's projects FIS2014-52837-P and FIS2016-78859-P (AEI/FEDER). This research has made use of data, software and/or web tools obtained from the Gravitational Wave Open Science Center (https://www.gw-openscience.org), a service of LIGO Laboratory, the LIGO Scientific Collaboration and the Virgo Collaboration. LIGO is funded by the U.S. National Science Foundation. Virgo is funded by the French Centre National de Recherche Scientifique (CNRS), the Italian Istituto Nazionale della Fisica Nucleare (INFN) and the Dutch Nikhef, with contributions by Polish and Hungarian institutes.

\section*{References}
%\bibliography{bibliography} 
%\bibliographystyle{ieeetr}

\normalsize
\newpage
\begin{appendices}
\section{Angular Momentum Emission}
\label{AngularMomentum}
From the free Lagrangian for a U(1) gauge boson field, we find the Noether charges corresponding to spatial rotations (setting $A^0=0$):

\begin{equation}
    J^i = \int \diff ^3 x
    \langle-\epsilon^{ikl}\dot A^m
    x^k \partial^l A^m
    + \epsilon^{ikl} A^k
    \dot A^l\rangle,
\end{equation}

\noindent
where the averaging is to remove any gauge ambiguity. Realising that the volume element is $r^2 \diff\Omega \diff r$ and that $\diff r=\diff t$ for radially outgoing radiation, we find the angular momentum carried by a radiation field at a shell:

\begin{equation}
    \dot J^i = r^2\int \diff \Omega 
    \langle-\epsilon^{ikl}\dot A^m
    x^k \partial^l A^m
    + \epsilon^{ikl} A^k
    \dot A^l\rangle,
\end{equation}

\noindent
where we identify the first term as representing orbital angular momentum, and the second spin.

For the spin contribution, a difference from before is realising that after we have done a passive rotation on the dipole to express it along a $z$-axis, we need to do another, similar rotation on the free index of the term before doing the solid angle integral. Then:

\begin{equation}
    \dot S^i = \frac{\epsilon^{3kl}}{4\pi}\int \diff \Omega\,
    \begin{pmatrix}
    \sin\phi\sin\theta \\ \cos\phi\sin\theta \\ \cos\theta \end{pmatrix}^i
    \langle \dot Q'^k\ddot Q'^l\rangle \propto \delta^{i3},
\end{equation}

\noindent
which makes sense because of the axisymmetry of Keplerian motion.

After evaluating and averaging over one orbit, $\langle f\rangle = \frac{1}{T}\int_0^T \diff t\,f$, we find:

\begin{equation}
    \dot S^{z}
    = \frac{\alpha^2\beta^2\omega_0}{3}.
\end{equation}

For the orbital angular momentum term, we find it useful with a slightly different approach. Following \cite{maggiore_gravitational_2007}, we picture a general direction dipole as given by

\begin{equation}
    Q'^i(\mathbf{n}) = P^i_{\,\,j}(\mathbf{n})
    Q^j(t-r),\quad
    P_{ij} = \delta_{ij}-n_i n_j,
\end{equation}

\noindent
where $P$ is a projection operator that enforces the transversal gauge in the $\mathbf{n}$-direction.

We have $\partial^i f(t-r) = -n^i \dot f$, so for the orbital angular momentum term, we find:

\begin{equation}
    \dot L = -\frac{\epsilon^{ikl}}{4\pi}r\langle \ddot Q^a \dot Q^d\rangle
    \int \diff \Omega\, P^{ba} n^k \partial^l P^{bd}.
\end{equation}

\noindent
We use the relations that we for brevity will not prove here:

\begin{equation}
    \partial^l P^{bd}
    = - \frac{1}{r}P^{lm}(
    \delta^{dm}n^b + \delta^{bm}n^d),\quad
    \int \diff \Omega\, n^d n^k = \frac{4}{3}\pi \delta^{dk},
\end{equation}

\noindent
which allow us to find the emission:

\begin{equation}
    \dot L = \frac{\alpha^2\beta^2\omega_0}{3},
\end{equation}

\noindent
so that the total angular momentum emission reads:

\begin{equation}
    \dot J = 
    \frac{2}{3}\alpha^2\beta^2\omega_0 .
\end{equation}

\section{Charge Quadrupole}
\label{chargequadrupole}

In the case where the charge to mass ratio difference is small, the charge quadrupole might be of significance. We consider equation \eqref{multipole}, and find the second term in terms of our multipoles, using current conservation $\partial_\mu j^\mu = 0$, yielding

\begin{align}&
    \dot Q^{ij} = 2 S^{(i,j)}\\
    \implies & \dot S^{i,j} = \frac{\ddot Q^{ij}}{2} + \dot S^{[i,j]},
\end{align}

\noindent
where the second term is the current dipole and is zero for a circular orbit, which we can see by, using equations \eqref{densities}-\eqref{cartesian},

\begin{align}
    S^{[i,j]} \propto \dot R^{[i} R^{j]} \propto \begin{pmatrix}
    0 & -1 \\ 1 & 0 \end{pmatrix}^{ij\in x,y}= \text{const}.
\end{align}

We want to derive the power for some charge dipole and quadrupole then. We use, as in appendix \ref{AngularMomentum}, the projection operator to allow for a general orientation for the multipoles, so that

\begin{align}
    - \frac{\diff ^2 E}{r^2 \diff t \diff \Omega} = P_{ij}\langle \dot A^i \dot A^j\rangle
    = \frac{P^{ij}}{4\pi r^2}\left\langle \ddot Q^i \ddot Q^j + \frac{n^k n^l}{4}  \dddot Q^{ik}\dddot Q^{jl}
    + n^{k}\dddot Q^{k(i} \ddot Q^{j)} \right\rangle,
\end{align}

\noindent
where we have used $P^{ij}P_{jk} = P^i_k$, and the last term is the dipole-quadrupole interaction and is automatically zero because the resulting angular integral will be over an odd number of $n^i$.

There not being any interaction, we may use the dipole power we have already found and work with the quadrupole power independently. We note however that in the event where the dipole is not vanishing, there would be an $\mathcal{O}(v^2)$ correction to the dipole coming from \spelloyvind{e.g.} the Biot-Savart interaction of the charges. We do not consider this here, so the following will only be the complete emission to charge quadrupole order when the Biot-Savart term is higher order relative to the charge quadrupole due to the dipole being suppressed from small charge to mass ratio difference. For the complete 1PN Lagrangian, see \cite{khalil_hairy_2018}. For the charge quadrupole term, taking the angular integral, we find

\begin{align}&
    \int \diff ^3 x\,\, n^i n^j n^k n^l = \frac{4\pi}{15}\left(
    \delta^{ij}\delta^{kl} + \delta^{ik}\delta^{jl} + \delta^{il}\delta^{jk}
    \right)\\
    \implies &
    P_{\text{EM}}^{\text{quad}} = \frac{1}{20} \left\langle \dddot Q^{ij}\dddot Q_{ij}
    - \frac{1}{3}\dddot Q^i_{\,\,j} \dddot Q^j_{\,\,j} \right\rangle .
\end{align}

\noindent
The second term effectively removes the trace of the quadrupole and it can be shown to be equivalent to removing it from the start, before all the derivatives and the contraction, and so for the traceless quadrupole, we have

\begin{equation}
    P = \frac{1}{20}\left\langle\dddot Q^{ij}\dddot Q^{ij}\right\rangle,
\end{equation}

\noindent
which is completely the same as for the reduced mass quadrupole radiation, except suppressed by a factor of 4.

We then look explicitly at the charge quadrupole and find

\begin{align}&
    \int \diff ^3 x\,\, j^t x^i x^j = \mu^2 \left(
    \sigma_1 / m_1 + \sigma_2 / m_2 \right) R^i R^j \\&
    = \mu (\sigma_1/ m_1 + \sigma_2 / m_2) M^{ij},
\end{align}

\noindent
where $M^{ij}$ is the mass quadrupole with trace from the GW multipole expansion.

We can then tell that the combined quadrupole power is

\begin{align}&
    P_{\text{tot}}^{\text{quad}} =  P_{GW}\Xi, \\&
    \Xi = 1 + \mu^2 (\sigma_1 /m_1 + \sigma_2 /m_2)^2/4 .
\end{align}

\section{Matched Filtering Numerics}
As we mentioned in section \ref{numericsInText}, our numerical scheme has been to construct a discrete grid of uncharged templates that we match charged signals onto. For each, we evaluate the SNR, according to equation (7.47) found in chapter 7 of \cite{maggiore_gravitational_2007}:

\begin{align}
    \text{SNR} = \frac{(u|h)}{\sqrt{(u|u)}}, \label{SNRintegrand}
\end{align}

\noindent
where the inner product is defined as the integral over frequencies of the product of their complex Fourier transforms (marked with tilde)

\begin{equation}
    (a|b) = 2 \text{Re} \int_{-\infty}^{\infty} df\,  \frac{\tilde a^* \,\tilde b}{S_n(f)} ,
\end{equation}

\noindent
and asterisk means that we take the complex conjugate, and $S_n$ is the noise spectral density explained in the reference, which we construct out of the aLIGO sensitivity curve that we found in \cite{arun_parameter_2005}. We then evaluate the SNR above for each point in our grid and select whichever template yields the highest number as our assumed source; Its parameters would be what is inferred for the source's parameters. However, in this simple description of our method hides some subtleties that we would like to clarify here.

\subsection{Units}

First, as is usual in numerics to deal with machine precision, we scale quantities as best we can so as to get the scale of our calculations to be about order $\mathcal{O}(1)$. We do this by using units $G = 1/4\pi\epsilon_0 = c = 1$ as is done otherwise in this article, but set additionally the solar mass $M_\odot =1$. This renders everything dimensionless and can be thought of as setting the energy scale. We have the following rules

\begin{align}&
    \text{To time, multiply } \frac{G M_\odot}{c^3} ,\\&
    \text{To mass, multiply } M_\odot ,\\&
    \text{To length, multiply } \frac{G M_\odot}{c^2},
\end{align}

\noindent
that may be used at any time to convert back to normal units.

\begin{table} 
    \centering
\begin{tabular}[centre]{c|c|c}
    Symbol & Value & Description \\\hline
    $\tau_n$ & $0.05$ s & normalisation time of templates and signal \\
    $S_0 \sim N^2$ & $1$ & spectral noise amplitude normalisation \\
    $f_{\text{min}}$ & $30$ Hz & lower frequency band of detector \\
    $f_{\text{max}}$ & $2$ kHz & upper frequency band of detector \\
    $f_0$ & $215$ Hz & frequency normalisation \\
    $f_s$ & $4096$ Hz & sampling rate \\
    $\diff \mathcal{M}$ & $0.02\,M_\odot$ & step size of grid for chirp mass \\
    $\diff \left[\mu\left(\Delta\sigma\right)^2\right]$ & $0.1 \, M_\odot$ & step size of charge parameter for signal
\end{tabular}
\caption{General quantities used in code for finding the chirp mass bias for charged signals. $N\sim$ amplitude of noise/signal in time-series.}
\label{table:quantitiesnumerics}
\end{table}
\begin{table} 
    \centering
    \begin{tabular}[centre]{c|c|c|c}
        Symbol & GW151226 & GW150914 & Figure \ref{fig:chirp}  \\ \hline
        d$\Phi_0$ & $\pi/2048 $& $\pi/360$ & $\pi/3000 $\\ $(\Phi_0|_{\text{start}}, \Phi_0|_{\text{stop}}) $&
        $(0,-\pi/50)$ &$ (0, -\pi/5) $& $(0, -2\pi/3)$
    \end{tabular}
    \caption{Simulation specific quantities used in code for finding chirp mass bias for charged signals. Chosen after some preliminary, coarse-grained matchings.}
    \label{table:quantities_simulationsspecific}
\end{table}

\newpage
\subsection{Template Grid and Signal Construction}

We construct a time-series mimicking the strain that would have been observed at a detector for a given signal in the stationary noise. The result looks like in figure \ref{fig:strain}. The noise and signal are simply added together at each point in time.

\subsubsection*{Noise Production:}

To produce the noise, we make use of the phenomenological aLIGO noise model taken from \cite{arun_parameter_2005}. It is non-trivial to draw stochastic variables out of a distribution defined by this spectral noise amplitude using pseudorandom numbers generated in Python. We construct the noise in frequency space, using the relation

\begin{align}
    n_i = \frac{\text{N}(0,1) + i\,\text{N}(0,1)}{\sqrt{2}}\sqrt{N_{\text{points}}\,\diff t\, S_{n}(f_i)} ,\label{noisedraw}
\end{align}

\noindent
where N$(0,1)$ is a draw from the normal distribution of mean 0 and standard deviation 1 while $S_n$ is the noise spectral density. This is the complex Fourier transform where there is an imaginary part, but because we know that the time-series should come out real, the coefficients of the negative frequencies have to be the conjugate of the respective positive ones. Taking care to convert our frequency space vector to the correct format for our fast Fourier transform function, we finally find the noise vector as a time-series. To make sure our noise production comes out right in relation to our spectral noise density, we find the periodogram and compare them, finding a perfect match.

\subsubsection*{Waveform Production:} 

For the other component of the strain data, we want to construct a signal. We base it on the equation given in \eqref{template}, and normalise it to unity at some small time before the merger, $\tau_n$, given in table \ref{table:quantitiesnumerics}. This ensures that the waveform\spelloyvind{'s amplitude} is \spelloyvind{similar} to \spelloyvind{that of} the noise floor, because we also have set $S_0=1$. The actual value of $S_0$ is $10^{-49}$Hz. The waveform is produced as a time series of $\,2 f_s \tau_0\,$ points, where $f_s$ and $\tau_0$ are given in table \ref{table:quantitiesnumerics} and \ref{tab:parameterchanges} respectively, extending from $t=-\tau_0$ to $t=\tau_0$ and the merger happens at $t=0$. To each sample point we add noise from the fast Fourier transform of \eqref{noisedraw} for the signal, while keeping the templates clean.

\subsubsection*{Grid Construction:}

We construct our grid of uncharged templates all with the same initial time until coalescence, varying them over chirp masses and initial phases. We write our code so that we can extend the grid to be varied over charges as well, and find when we do this that the correct template is selected for. Initially, we use somewhat large matching intervals, but narrow them once it is clear which part of them the signal is projected onto, and increase the grid point density in these intervals instead. As a check of whether the densities are high enough, we check that we recover the correct signal when it is included in our template bank, we check that we would recover an uncharged signal, and that both the averaged estimated parameters evolve somewhat smoothly with charge after a number of realisations. We also check that slightly increasing our grid densities does not change our estimations significantly. Our final step sizes in the different quantities of the grid are shown in table \ref{table:quantitiesnumerics} and \ref{table:quantities_simulationsspecific}.

\subsection{Matching Procedure}

Our matching function is based on a tutorial found on the website of the LIGO Open Science Center \cite{vallisneri_ligo_2015}. Instead of doing the inverse fast Fourier transform of the kernel of the signal-to-noise ratio integrand \eqref{SNRintegrand} to find the SNR as a time-series where the template and strain are arbitrarily displaced, us knowing exactly where the signal in the strain is hidden, we simply perform the integration for perfect overlap between strain and template. We thus assume that this is what the time-series would have selected for anyway, which we found to be true in our initial testing using the time-series approach. This saves us the computational effort of one Fourier transform each matching. However, we do vary the templates over initial phases to cover a possible bias arising there.

\subsection{Monte Carlo}

Once we have matched a strain over all the templates in the grid, the template with the highest SNR is selected for and its parameters recorded as the inferred ones for the source. We repeat this for different charges with a charge parameter step size given in table \ref{table:quantitiesnumerics} up until the perturbation quantity is of order $\sim 0.1$ so that the linearisation falls apart. The noise being stochastic, to find convergence in our results, we repeat this whole matching a different number of times for the different simulations and take the average of the inferred source parameters, which we, for the chirp mass, for the long inspiral scenario, then plot in figure \ref{fig:chirp} together with the SNR relative to the SNR that we would have had if we had the correct charged template contained in the template bank.
\newline

\noindent
\textbf{Code: } The code used\spelloyvind{ in this work} may be found at\newline {\small \url{https://github.com/oyvach/matched-filtering-mock-bias}}.

\end{appendices}

\end{document}